\documentclass[pre,aps,reprint,noshowpacs,superscriptaddress,floatfix,letterpaper,longbibliography]{revtex4-2}
\usepackage{amsmath,amssymb,amsbsy,amsfonts,amsthm,bbm,bm,mathtools,mathrsfs}
\usepackage{color}
\usepackage{physics}
\usepackage{xfrac}
\usepackage[dvipsnames]{xcolor}
\definecolor{LapisLazuli}{RGB}{47, 102, 169}
\usepackage[colorlinks=true,citecolor=LapisLazuli,linkcolor=LapisLazuli,urlcolor=LapisLazuli]{hyperref}
\usepackage{empheq}
\usepackage{pgfplots}
\usepackage{stackengine}
\usepackage{relsize}
\usepackage[inline]{enumitem}
\usepackage[normalem]{ulem} 
\usepackage{comment}
\usepackage{import}
\usepackage{soul}
\usepackage[english]{babel}
\UseRawInputEncoding
\usepackage{pgfplots}
\pgfplotsset{compat = newest}
\usetikzlibrary{arrows,intersections}
\usepackage{tikz-3dplot}
\usepackage{tikz}

\usepackage[mathscr]{euscript}

\newcommand{\CR}{Cram{\'e}r-Rao\,}

\usepackage{calligra}

\DeclareMathAlphabet{\mathcalligra}{T1}{calligra}{m}{n}
\DeclareFontShape{T1}{calligra}{m}{n}{<->s*[2.2]callig15}{}

\newcommand{\speedtime}{\tau_{\!\scriptscriptstyle \mathcal{O}}}
\newcommand{\speedtimep}{\tau_{\!\scriptscriptstyle \mathcal{O},+}}
\newcommand{\speedtimem}{\tau_{\!\scriptscriptstyle \mathcal{O},-}}
\newcommand{\speedtimepm}{\tau_{\!\scriptscriptstyle \mathcal{O},\pm}}
\newcommand{\poisson}{\!\scriptscriptstyle P}
\newcommand{\bx}{\boldsymbol{x}}
\newcommand{\stability}{{\boldsymbol{A}}}

\newcommand{\ex}{{\boldsymbol{x}}}
\newcommand{\obs}{{\boldsymbol{O}}}
\newcommand{\yu}{{\boldsymbol{u}}}

\newcommand{\brho}{{\boldsymbol{\varrho}}}
\newcommand{\bxi}{{\boldsymbol{\xi}}}
\newcommand{\logder}{{\boldsymbol{L}}}
\newcommand{\bX}{{\boldsymbol{X}}}
\newcommand{\bY}{{\boldsymbol{Y}}}

\usepackage{pgfplots}
\pgfplotsset{compat = newest}
\usetikzlibrary{arrows,intersections}
\usepackage{tikz-3dplot}
\usepackage{tikz}
\makeatletter
\def\maketitle{
	\@author@finish
	\title@column\titleblock@produce
	\suppressfloats[t]}
\makeatother
\begin{document}
	
\def\xlist{4}
\def\ylist{4}

\title{Speed limits on classical chaos}

\author{Swetamber~Das}
\author{Jason~R.~Green}
\email[]{jason.green@umb.edu}
\affiliation{Department of Chemistry,\
	University of Massachusetts Boston,\
	Boston, MA 02125
}
\affiliation{Department of Physics,\
	University of Massachusetts Boston,\
	Boston, MA 02125
}

\begin{abstract}

Uncertainty in the initial conditions of dynamical systems can cause exponentially fast divergence of trajectories, a signature of deterministic chaos.
Here, we derive a classical uncertainty relation that sets a speed limit on the rates of local observables underlying this behavior. For systems with a time-invariant stability matrix, this general speed limit simplifies to classical analogues of the Mandelstam-Tamm versions of the time-energy uncertainty relation.
This classical bound derives from our definition of Fisher information in terms of Lyapunov vectors on tangent space, analogous to the quantum Fisher information defined in terms of wavevectors on Hilbert space.
This information measures fluctuations in local stability of the state space and sets a lower bound on the time of classical, dynamical systems to evolve between two distinguishable states.
The bounds it sets apply to systems that are open or closed, conservative or dissipative, actively driven or passively evolving, and directly connect the geometries of phase space and information.

\end{abstract}

\maketitle

Quantum speed limits are fundamental constraints on the time evolution of quantum mechanical systems and their observables~\cite{deffnerQuantumSpeedLimits2017}.
A milestone in their development is the Mandelstam-Tamm versions of the time-energy uncertainty relation, which sets a speed limit on the observables of unitary quantum dynamics~\cite{mandelstamUncertaintyRelationEnergy1991}.
This and other bounds have been extended~\cite{buschTimeEnergyUncertainty2008} to open quantum systems~\cite{taddeiQuantumSpeedLimit2013,delcampoQuantumSpeedLimits2013,deffnerQuantumSpeedLimit2013,garcia-pintosQuantumSpeedLimits2019} and applied to many-body dynamics~\cite{fogartyOrthogonalityCatastropheConsequence2020,delcampoProbingQuantumSpeed2021}. They have also been connected to parameter estimation~\cite{braunsteinGeneralizedUncertaintyRelations1996,giovannettiAdvancesQuantumMetrology2011,beauNonlinearQuantumMetrology2017,sidhuGeometricPerspectiveQuantum2020} and information theory~\cite{jingFundamentalSpeedLimits2016,deffnerQuantumSpeedLimits2020,piresBoundingGeneralizedRelative2021} where they quantify the inherent limits on measurements of dynamical quantities~\cite{margolusMaximumSpeedDynamical1998,lloydUltimatePhysicalLimits2000}.
It was recently discovered that there are bounds on the evolution of classical systems, the earliest of which largely rely on the Hilbert space of the Liouville equation~\cite{shanahanQuantumSpeedLimits2018,okuyamaQuantumSpeedLimit2018}.
For purely stochastic dynamics, there is now a growing number of thermodynamic speed limits~\cite{shiraishiSpeedLimitClassical2018,hasegawaUncertaintyRelationsStochastic2019,itoStochasticTimeEvolution2020,nicholsonNonequilibriumUncertaintyPrinciple2018,nicholsonTimeInformationUncertainty2020} on the flux of energy and entropy between a system and external reservoirs. Included among them is a stochastic thermodynamic speed limit~\cite{nicholsonTimeInformationUncertainty2020} that, when combined with the Mandelstam-Tamm bound, %
gives a more general speed limit on the observables of open quantum systems~\cite{garcia-pintosUnifyingQuantumClassical2021}.
Despite this progress, all the currently known classical speed limits are on statistical dynamics, leaving open the question of whether there are speed limits on the underlying physical dynamics, dynamics that often exhibit deterministic chaos. We address this question here.

While many deterministic systems do not have stochastic fluctuations, they can be characterized by ``uncertainty'' associated with their evolution that originates from small disturbances in their initial conditions.
The divergence of initially close phase space trajectories~\cite{gaspardChaosScatteringStatistical1998}, with the local rates of divergence providing an intrinsic timescale for the exploration of state space, is a characteristic of deterministic chaos.
Deterministic chaos appears in the behavior of many classical systems evolving under a strongly nonlinear dynamics.
Measures of chaos have given insights into the physical mechanisms of the jamming transition in granular materials~\cite{baniganChaoticDynamicsJamming2013}, self-organizing systems~\cite{greenRelationshipDynamicalEntropy2013}, evaporating collections of nuclei, equilibrium and nonequilibrium fluids~\cite{evansStatisticalMechanicsNonequilibrium, bosettiWhatDoesDynamical2014, dasSelfaveragingFluctuationsChaoticity2017}, and critical phenomena~\cite{dasCriticalFluctuationsSlowing2019}.
This feature of dynamical systems, and the growing connections between dynamical systems theory and nonequilibrium statistical mechanics~\cite{dorfmanIntroductionChaosNonequilibrium1999,gaspardChaosScatteringStatistical1998}, suggest the possibility of classical speed limits on the intrinsic timescales of dynamical instability that underlie deterministic chaos.

In this Letter, we derive classical bounds on the observables and the state space of deterministic systems that exactly parallel the Mandelstam-Tamm form of the time-energy uncertainty relation in quantum mechanics.
Mandelstam and Tamm~\cite{mandelstamUncertaintyRelationEnergy1991} considered isolated quantum systems evolving unitarily,
proving that the rate of change of the expectation value $\langle \hat O \rangle$ of an arbitrary quantum observable, $\hat O$, is bounded,
$\vert d \langle \hat O \rangle/dt \vert \leq 2 \Delta \hat O \Delta \hat H$,
by the standard deviations of the observable and the Hamiltonian, $\hat H$.
Perhaps more well known is their result that the minimum time $\tau^{\perp}$ for a system to evolve between two orthogonal states satisfies $\tau^{\perp} \geq \pi /(2 \Delta \hat H)$. Here, we derive purely classical analogues of these bounds for dynamics that are not statistical in nature.
For deterministic, physical dynamics, we define an intrinsic timescale for the mean of a given dynamical observable to change by the value of one initial standard deviation and the first definition of the Fisher information for these dynamics.
These enable us to derive speed limits on the underlying dynamics of any classical system, be it open or closed, continuous or many-body, dissipative or conservative, passively evolving or actively driven.

\textit{Heuristic argument.---} Intuition for these speed limits comes from a simple heuristic argument. As in quantum speed limits, a classical speed limit requires the identification of an intrinsic speed for the dynamics.
A natural choice for the speed is in the local stability that causes the convergence/divergence of classical trajectories.
Take two points, initially arbitrarily close, separated by a distance $\|\delta\bx(t_0)\|=\|\bx(t_0)-\bx^{\prime}(t_0)\|$.
When the dynamics are sensitive to initial conditions, the distance grows to $\|\delta\bx(t)\|$ at a rate $r(t)$ over a time $\Delta t= t-t_0$; the largest finite-time Lyapunov exponent, $\lambda$, measures the exponential rate at which the distance between two trajectories diverges (or converges) in state space, $e^{\lambda \Delta t}$~\cite{pikovskyLyapunovExponentsTool2016}.
With this Lyapunov exponent, a plausible speed limit would be the time for the distance to grow by $e$:
$\|\delta\bx(t)\|/\|\delta\bx(t_0)\| \approx e$ or $\ln\|\delta\bx(t)\|-\ln\|\delta\bx(t_0)\| = \lambda\, \Delta t\geq 1$. 
Loosely speaking, the divergence of initially close trajectories would determine the timescale $\Delta t$ on which the behavior of the system is predictable and beyond which it is ``chaotic''.
The time it takes for the distance
between two nearby phase points %
to increase by exactly a factor of $e$, known as the Lyapunov time, $\Delta t_{\scriptscriptstyle\operatorname{L}}$, would saturate this heuristic bound $\lambda \Delta t_{\scriptscriptstyle\operatorname{L}} \approx 1$. These ideas prompt a more careful derivation of speed limits from the local dynamical (in)stability.

\textit{Dynamics of the classical density matrix.---} To derive a speed limit more rigorously, and for other observables, we instead start from the linearized dynamics.
Take a dynamical system, $\dot{\ex} = \boldsymbol{F}[\ex(t)]$, where $\ex$ represents a point $\ex(t) := [x^1(t),x^2(t),\ldots,x^n(t)]^\top$ in the $n$-dimensional state space.
Infinitesimal perturbations $\ket{\delta \ex(t)}$ in the tangent space represent uncertainty about the initial condition; these vectors stretch, contract, and rotate over time under the linearized dynamics,
\begin{equation}
  d_t\ket{\delta \ex(t)} = \stability[\ex (t)]\ket{\delta \ex(t)},
\label{equ:EOM_per_ket2}
\end{equation}
with the stability matrix, $\stability:=\stability[\ex(t)] = \grad\boldsymbol{F}$ having the elements
$(\stability)^i_{j}=\partial \dot{x}^i(t)/\partial x^j(t)$.
Dirac's notation here represents a finite-dimensional column (row) vector with the ket (bra): $\ket{\delta\ex(t)}:= [\delta x^1(t), \delta x^2(t), \ldots, \delta x^n(t)]^\top\in T\mathcal{M}$.
These linearized dynamics are an established approach to analyze the stability of nonlinear dynamical systems~\cite{pikovskyLyapunovExponentsTool2016}.
Solving the equation of motion,
\begin{align}\label{equ:per_ket}
\ket{\delta \ex(t)}
&= \mathcal{T}_+e^{\int_{t_0}^{t+t_0}\stability(t')\,dt'}\ket{\delta \ex(t_0)},
\end{align}
gives the perturbation vector at a time $t$ in terms of the propagator with time ordering through the operator $\mathcal{T}_+$.
This evolution operator is reminiscent of the Dyson series and interaction picture in quantum mechanics~\cite{joachainQuantumCollisionTheory1983}.

Unlike the quantum Hamiltonian, which is Hermitian, the stability matrix $\stability$ is generally not symmetric and, so, the tangent space evolution operator in Eq.~\ref{equ:per_ket} is generally not unitary.
However, there is a norm-preserving operator for the tangent space dynamics of a \textit{unit} perturbation vector. A normalized, classical density matrix  in terms of a unit perturbation $\ket{\delta \yu(t)} = \ket{\delta\bx(t)/\|\delta \bx(t)\|}$ is a projection operator, $\brho(t) = \dyad{\delta \yu(t)}{\delta \yu(t)}$, with the properties one expects of a pure state~\cite{dasDensityMatrixFormulation2021}.
Its time evolution is governed by an equation of motion,
\begin{align}\label{eq:EOM_rho}
	d_t\brho = \{\stability_+,\brho\} + [\stability_-,\brho] - 2\langle\stability_+\rangle\brho,
\end{align}
akin to the von Neumann equation in quantum dynamics. Here, $\langle\stability_+\rangle=\Tr(\stability_+\brho)$ and $\stability_\pm$ represent the symmetric and anti-symmetric parts of $\stability$ appearing in the anti-commutator $\{\bX,\bY\} = \bX\bY + \bY\bX$ and the commutator $[\bX,\bY] = \bX\bY - \bY\bX$, respectively.
This norm-preserving dynamics holds regardless of whether the dynamical system is Hamiltonian or dissipative and enables a generalization of Liouville's theorem and Liouville's equation on phase space volumes~\cite{dasDensityMatrixFormulation2021}.

\textit{Equation of motion for observables.---} Natural observables are expectation values with respect to this density matrix, $\brho$, on the deterministic tangent space dynamics. For example, the instantaneous Lyapunov exponent or \textit{local} stretching rate for the linearized dynamics is, $r=r(t):=r[\ex(t)] = d_t\ln\|\delta \yu(t)\|=\langle\stability_+\rangle=\Tr(\stability_+\brho)$.
This local rate is related to the finite-time Lyapunov exponent,
\begin{equation}
\lambda(t) = |t|^{-1}\int_{t_0}^{t+t_0}r[\ex(t^{\prime})]\,dt^{\prime} = \frac{1}{2t}\ln \Tr(\bxi\brho), %
\end{equation}
measuring phase space (in)stability.
Here, $\bxi=\dyad{\delta \ex(t)}{\delta \ex(t)}$ is the unnormalized projector that can be normalized to construct a ``pure state'' from a single perturbation vector $\brho = \dyad{\delta \yu}{\delta \yu}$
\footnote{Our results also hold for \emph{maximally} mixed normalized states, $\brho'(t) = k^{-1}\sum_{i=1}^{k}\brho_i(t)$, which also evolve according to Eq.~\ref{eq:EOM_rho}.
In this case, expectation values are to be computed with respect to $\brho'$.}.

Lyapunov exponents have been connected to thermodynamic properties, such as energy dissipation and entropy production, and transport properties, such as the diffusion and viscosity coefficients~\cite{Evans1990,GasNico1990,CohenRondoni1998,Ruelle1999,greenRelationshipDynamicalEntropy2013,Qian2019EntropyPI,Caruso2020}.
These connections can derive from the sum of instantaneous Lyapunov exponents, which determines the \emph{local} phase space contraction rate $\Lambda$ at a given phase point: $\Lambda = \sum_{j=1}^n r_j = \sum_{j=1}^n\Tr(\stability_+\brho_j)$, where the $\brho_j$ are the projections from a complete set of linearly independent tangent vectors at the phase point in an $n$-dimensional phase space.
When averaged over a stationary distribution, the negative phase space volume contraction rate is the Gibbs entropy production rate for Gaussian-thermostatted, time-reversible dynamical systems~\cite{Daems1999,Patra2016, Ramshaw2017,Qian2019EntropyPI}.
Both $\Lambda$ and the entropy production have upper and lower bounds in a 2D Lorentz gas subject to a Gaussian thermostat~\cite{dasDensityMatrixFormulation2021}.
Given these connection between dynamical quantities and physical observables, we will first establish speed limits on Lyapunov exponents.
(Our results that follow from this point may be extended to speed limits on physical observables through $\Lambda$.)

From the dynamics of the state, we derive the time evolution of the moments $\langle \obs \rangle$ of an observable $\obs$ (Supplemental Material, SM~I):
\begin{align}\label{eq:EOM_mean_obs}
d_t\langle\obs\rangle &= \operatorname{cov}(\obs,2\stability^\top) + \left\langle d_t\obs\right\rangle,
\end{align}
the tangent space analogue of Ehrenfest's theorem (Heisenberg's equation) for quantum mechanical observables~\cite{messiahQuantumMechanics1999} with an additional anticommutator term. The covariance,
$\operatorname{cov}(\boldsymbol X,\boldsymbol Y) = \langle \boldsymbol X \boldsymbol Y^\top\rangle -\langle \boldsymbol X \rangle \langle \boldsymbol Y^\top \rangle$, is composed of two pieces: the mean anticommutator, $\operatorname{cov}(\obs,2\stability_+) = \langle\{\obs,\stability_+\} \rangle - 2\langle \stability_+\rangle\langle\obs\rangle$, and the mean commutator, $\operatorname{cov}(\obs,2\stability_-)=\langle [\obs,\stability_-] \rangle$.
Both are symmetric matrices and themselves tangent space observables.
Equation~\ref{eq:EOM_mean_obs} is also similar in mathematical form to the equation of motion for stochastic thermodynamic observables~\cite{nicholsonTimeInformationUncertainty2020} and Price's equation in population biology~\cite{priceSelectionCovariance1970}.
If the dynamics are Hamiltonian, Eq.~\ref{eq:EOM_mean_obs} can be expressed, $\{H,\langle\obs\rangle\}_{\poisson} = \operatorname{cov}(\obs,2\stability^\top) + \left\langle d_t\obs\right\rangle$,
in terms of the Hamiltonian of the system $H$ and the Poisson bracket $\{.\}_{\poisson}$.
If the observables $\obs$ and $\brho$ commute (i.e.,
if they are share the same set of eigenbasis), the covariance vanishes, $\operatorname{cov}(\obs,2\stability^\top) = 0$, and Eq.~\ref{eq:EOM_mean_obs} reduces to $d_t\langle \obs \rangle= \left\langle d_t\obs \right\rangle$(SM~II).
With the equation of motion for the moments of observables, we can derive classical uncertainty relations that set limits on the intrinsic speed at which they evolve --- limits set by the Fisher information.

\textit{Tangent-space Fisher information.---}
Another observable of interest is the Fisher information, which is a fundamental ingredient in optimal measurements of random variables, setting a lower bound on the variance of unbiased estimators of parameters through the \CR information inequality.
It is also an intrinsic speed on the evolution of a system betweeen neighboring states~\cite{kimGeometricStructureGeodesic2016,flynnMeasuringDisorderIrreversible2014,*nicholsOrderDisorderIrreversible2015}.
However, it also has a geometric representation through the Fisher information matrix, a Riemannian metric on statistical manifolds~\cite{sidhuGeometricPerspectiveQuantum2020}.
Because of the importance of the Fisher information in parameter estimation, it is not immediately clear that it is relevant for physical dynamics. However, the density matrix representation of these dynamics allows us to overcome this conceptual challenge.

From the norm-preserving dynamics of the classical density matrix, $\brho$, we can define a Fisher information (matrix) on the perturbation vectors, $\ket{\delta \yu}$, in tangent space.
In terms of the deviation $\bar{\stability} = \stability - \langle \stability\rangle$, Eq.~\ref{eq:EOM_rho} becomes:
\begin{align}\label{eq:EOM_rho_logder}
  d_t\brho = \bar{\stability}\brho + \brho \bar{\stability}^\top.
\end{align}
This equation of motion for the density matrix defines a logarithmic derivative $\logder$ implicitly through $d_t\brho := \frac{1}{2}(\brho\logder + \logder^\top\brho)$~\cite{fujiwaraQuantumFisherMetric1995,tsangFundamentalQuantumLimit2011}, SM III.
Using $\logder$, the equation of motion for an observable $\obs$ in Eq.~\ref{eq:EOM_mean_obs} is: $d_t\langle\obs\rangle = \operatorname{cov}(\obs,\logder) + \left\langle d_t\obs\right\rangle$.
A natural definition of the tangent-space Fisher information for pure states in a basis-independent form,
\begin{align}\label{eq:def_FI}
\mathcal{I}_F = \Delta \logder^2 = \langle\logder \logder^\top\rangle= 4(\Delta \stability^\top)^2,
\end{align}
is as the expectation value of the Fisher information matrix $\logder\logder^{\top}$.
This Fisher information is also the variance of the total logarithmic derivative because the total logarithmic derivative $\logder = 2\bar{\stability}^\top = 2( \stability^\top - \langle \stability\rangle)$ has a mean $\langle \logder \rangle = 0$.
It is also the variance in local stability $(\Delta \stability^\top)^2 = \langle \stability^\top \stability\rangle - \langle \stability_+ \rangle^2$ for pure states.
As a point of comparison, the quantum Fisher information is the variance in the energy $\mathcal{\hat I}_F = 4 \Delta
\hat{\boldsymbol{H}}^2/\hbar^2$ for pure quantum states evolving under a unitary
dynamics~\cite{Helstrom1967}.
The parallels between the quantum Fisher information and the classical Fisher information on tangent space here suggests the potential for speed limits on classical (potentially non-stochastic) dynamics.

\textit{Time-information uncertainty relations.---} With the preceding groundwork, we can derive these speed limits on time-evolving tangent space observables for classical, deterministic dynamical systems, including those determining the degree of classical chaos.
Rearranging Eq.~\ref{eq:EOM_mean_obs} gives, $\dot{\mathcal{O}}:=\operatorname{cov}(\obs,\logder) = d_t\langle\obs\rangle - \left\langle d_t\obs\right\rangle$.
One measure of the variation in $\obs$ is the time it takes for the magnitude of this function $\mathcal{O} = \int\dot{\mathcal{O}}dt$ to have the value of one standard deviation $\Delta \obs$. If $\dot{\mathcal{O}}$ is constant, this time $\speedtime$ is approximately:
\begin{align}
|\mathcal{O}| = \left| \int_{t_0}^{t_0 + \speedtime}  \dot{\mathcal{O}}dt\right| \approx |\dot{\mathcal{O}}|\speedtime \approx \Delta \obs.
\end{align}
This observation motivates the definition of an intrinsic speed for the time evolution of $\obs$,
\begin{align}\label{eq:def_tau}
\frac{1}{\speedtime}:=\frac{|\dot{\mathcal{O}}|}{\Delta \obs} = \frac{|\operatorname{cov}(\obs,\logder)|}{\Delta \obs},
\end{align}
similar to definitions in quantum mechanics~\cite{messiahQuantumMechanics1999,garcia-pintosUnifyingQuantumClassical2021} and stochastic thermodynamics~\cite{nicholsonTimeInformationUncertainty2020,nicholsonThermodynamicSpeedLimits2021}.

With this intrinsic speed on an observable, $\obs$, our main result is a limit that comes from applying the Cauchy-Schwarz inequality to the covariance gives a classical uncertainty relation:
\begin{align}
 \operatorname{cov}(\obs,\logder)^2 \leq \Delta \obs^2\,\Delta \logder^2 = 4\Delta \obs^2 \Delta \stability^{\top 2}.
\end{align}
Using the tangent-space Fisher information, this upper bound immediately leads to the uncertainty relation:
\begin{align}\label{eq:tiur}
  \tau_{\mathcal{O}} \sqrt{\mathcal{I}_F} \geq 1 \quad \text{or} \quad \tau_{\mathcal{O}}\,\Delta \stability^\top \geq \frac{1}{2}.
\end{align}
The latter form is a classical analogue of the Mandelstam-Tamm uncertainty relation in quantum mechanics. Currently, the Mandelstam-Tamm result is often cast as a speed limit $\tau^{-1}_{\hat O} \leq \tau^{-1}_{\textrm{QSL}}$ by defining the speeds $\tau_{\hat O}^{-1} = \Delta \hat O^{-1}d\langle \hat O\rangle/dt$ and $\tau^{-1}_{\textrm{QSL}}=1/(2\Delta \hat H)$.
Here, the tangent space Fisher information is the intrinsic speed $\sqrt{\mathcal{I}_F} = \tau^{-1}$ that sets the limit on the speed $\speedtime^{-1}$ of the observable, $\speedtime^{-1}\leq \tau^{-1}$.
As numerical support, Figure~\ref{fig:speed_chaos}(a) shows that this bound is satisfied for a chaotic orbit of the Lorenz model. In the case of 2D Li\'enard systems~\cite{Grasman2005}, such as the van der Pol oscillator, this speed limit on the only non-zero, local stretching rate, $\langle \stability_+\rangle$ is also a bound on $\Lambda$ and the energy dissipation rate.
\begin{figure}[t]
	\centering
	\includegraphics[width=\linewidth=1.0\columnwidth]{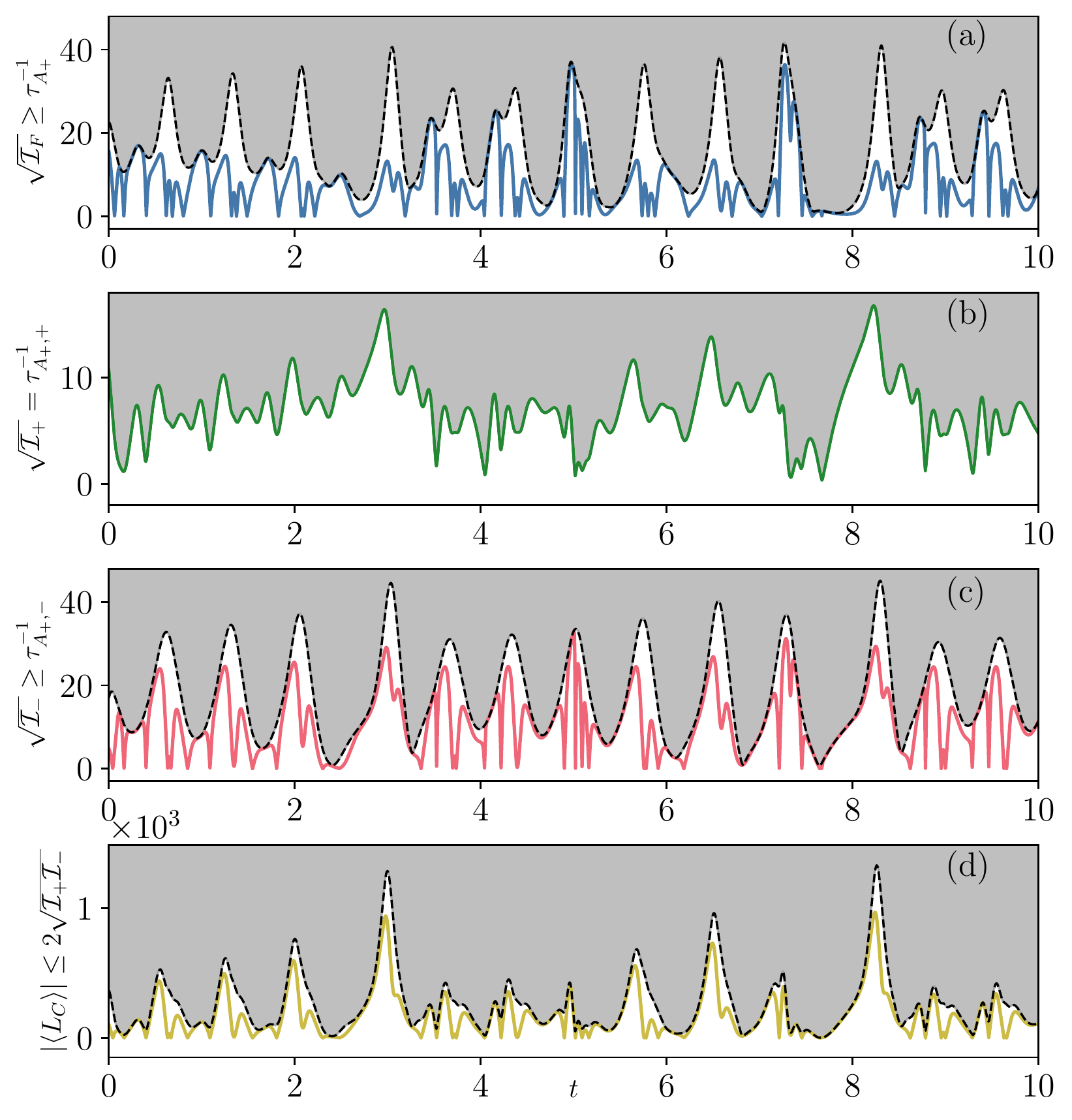}
	\caption{\label{fig:speed_chaos}\textit{Speed limit on chaos: the instantaneous Lyapunov exponent, $r(t) = \langle \stability_+\rangle$, for a chaotic orbit of the Lorenz model.---} (a) Square root of the tangent-space Fisher information (dashed black) $\sqrt{\mathcal{I}_F}$ upper bounds the speed $\tau_{\stability_+}^{-1}$ (solid blue). (b) For this observable, the speed $\tau_{\stability_+,+}^{-1}$ saturates the bound set by $\sqrt{\mathcal{I}_+}$ (solid green). (c) The speed $\tau_{\stability_+,-}$ (solid red) is bounded by $\sqrt{\mathcal{I}_-}$ (dashed black). (d) The mixed term $|\langle\logder_C\rangle|$ (solid yellow) is bounded by 2$\sqrt{\mathcal{I}_+\mathcal{I}_+}$ (dashed black) from above. Shaded regions mark speeds not accessible to the observable. The parameters of the Lorenz model are: $\sigma = 10$, $\beta = 8/3$, $\rho = 28$.}
\end{figure}

\textit{\CR bound.---} In quantum mechanics, the observable $\hat{\boldsymbol{Q}}$, the mean commutator $\langle[\hat{\boldsymbol{Q}}, \hat{\boldsymbol{H}}]\rangle$ takes the role of the covariance term in Eq.~\ref{eq:def_tau} and the term $\langle d_t \hat{\boldsymbol{Q}}\rangle$ in the Ehrenfest equation vanishes.
However, the corresponding term in classical dynamical systems, $\langle d_t\obs\rangle$ does not necessarily vanish.
It is nonzero for many dynamical systems because, unlike $\hat{\boldsymbol{H}}$, the stability matrix $\stability$ is commonly time dependent.
However, if the observable $\obs$ is time independent, the second term in Eq.~\ref{eq:EOM_mean_obs} vanishes, $d_t\langle \obs\rangle = \operatorname{cov}(\obs,\logder)$.
(A similar restriction on stochastic thermodynamic observables simplifies a more general bound~\cite{nicholsonTimeInformationUncertainty2020} for time-independent observables to bounds based on the \CR inequality~\cite{hasegawaUncertaintyRelationsStochastic2019,itoStochasticTimeEvolution2020}.)
Applying the Cauchy-Schwarz inequality, as before, gives,
 \begin{align}
  d_t\langle \obs \rangle \leq\Delta \obs\Delta \logder = \Delta \obs \sqrt{\mathcal{I}_F} = 2\Delta\obs \Delta \stability^\top,
 \end{align}
another classical analogue of the Mandelstam-Tamm uncertainty relation in quantum mechanics and the Cram\'er-Rao bound in classical statistics.

Following Mandelstam-Tamm~\cite{mandelstamUncertaintyRelationEnergy1991}, this speed limit can put a bound on evolution in the phase space.
Choosing the observable to be the projection of the initial state $\brho(t_0)=\dyad{\delta \yu(t_0)}{\delta \yu(t_0)}$, the time evolution of $\langle\brho(t_0)\rangle$ is lower bounded by $\langle \brho(t_0) \rangle \geq \cos^2(\Delta \stability^\top t)$ in the time interval $0\leq t\leq\pi/2\Delta \stability^\top$.
A similar result holds for the quantum mechanical mean density operator $\langle\hat{\rho}(t_0)\rangle$\footnote{The expression is $\langle \hat\rho(t_0) \rangle \geq \cos^2(\Delta \hat{\boldsymbol{H}} t/\hbar)$ for $0\leq t\leq\pi\hbar/2\Delta\hat{\boldsymbol{H}}$.
For details, see Ref.
~\cite{mandelstamUncertaintyRelationEnergy1991}.}.
The analogue of the time-energy uncertainty relation for classical systems follows:
\begin{equation}
\tau^\perp\Delta\stability^\top\geq \pi/2.
\end{equation}
for the time $\tau^\perp$ it takes for the initial state to evolve to orthogonal state.
Compared to Eq.~\ref{eq:tiur}, this bound holds for the comparatively few dynamical systems where $\stability$ is time independent; two examples are the harmonic oscillator and the model for Chua's circuit~\cite{chuaChuaCircuit101994}.
While most nonlinear systems will violate this bound, they will satisfy more general bound, Eq.~\ref{eq:tiur}, which holds regardless of the nature of dynamics.

\textit{Partitioned speed limits.---}
The global speed limit in Eq.~\ref{eq:tiur} is further divisible into speed limits on the symmetric and antisymmetric parts of the stability matrix.
In quantum estimation theory, $\logder$ is a symmetric operator~\cite{parisQuantumEstimationQuantum2009,sidhuGeometricPerspectiveQuantum2020}.
By contrast, our choice of the (total) logarithmic derivative here is dictated by Eq.~\ref{eq:EOM_rho_logder}.
However, it can be partitioned into its symmetric and antisymmetric parts,
\begin{align}
  \logder_+ = 2(\stability_+ - \langle \stability_+\rangle) \quad\text{and}\quad 	\logder_- = -2\stability_-,
\end{align}
with mean $\langle \logder_\pm\rangle = 0$ and variance $\Delta \logder_\pm^2 = \langle\logder^2_\pm \rangle$.
As a consequence, the Fisher information partitions into three components:
\begin{align}\label{eq:FI}
	\mathcal{I}_F = \mathcal{I}_+ + \mathcal{I}_- + \langle \logder_C\rangle.
\end{align}
The symmetric and antisymmetric parts are also variances $\mathcal{I}_\pm = \langle \logder_\pm^2\rangle = 4\Delta \stability_\pm^2$.
The third term is the commutator of logarithmic derivatives, $\logder_C = [\logder_-,\logder_+] = 4[\stability_+,\stability_-]$, a traceless symmetric matrix. (SM~V includes a geometrical interpretation.)
The \textit{mixed} term $\langle\logder_C\rangle$ is a covariance,
$\langle \logder_C\rangle =  4\,\operatorname{cov}(\stability_-,\stability_+)$, that follows the cosine law and is related to angle between tangent vectors $\logder_+\ket{\delta \yu}$ and $\logder_-\ket{\delta \yu}$, SM~IV.

Splitting the covariance term, $\operatorname{cov}(\obs,\logder) = \operatorname{cov}(\obs,\logder_+) + \operatorname{cov}(\obs,\logder_-)$, we identify two timescales:
$\speedtimepm^{-1}:= |\operatorname{cov}(\obs,\logder_\pm)|/\Delta\obs$.
This partitioning decouples the influence of the symmetric and anti-symmetric parts of the stability matrix on the evolution of the observable, operating on separate time scales $\speedtimep$ and $\speedtimem$. 
These timescales also have speed limits, again through the Cauchy-Schwarz inequality:
\begin{align}\label{eq:tiur_parts}
  \speedtimepm\sqrt{\mathcal{I}_\pm}\geq 1.
\end{align}
The Fisher information contributions $\mathcal{I}_{\pm}$ are related to the stretching/contraction and rotation of the vector $\ket{\delta \yu}$ (SM IV).
The former $\mathcal{I}_+$ is fixed by the instantaneous Lyapunov exponent and its magnitude sets the maximum threshold on, for example, the rate of stretching of a tangent vector.
The inequalities above are then speed limits on the development of chaos in any dynamical system.

Applying the triangle inequality to the covariance, $\operatorname{cov}(\obs,\logder)$, we find upper and lower speed limits on the timescale of the observable: $|\speedtimep^{-1} - \speedtimem ^{-1}| \leq \speedtime^{-1}  \leq \speedtimep^{-1} + \speedtimem ^{-1}$.
Combining this with Eq.~\ref{eq:FI}, suggests that the geometric mean of $\mathcal{I}_\pm$ sets an upper bound on $|\langle \logder_C\rangle|$ (SM V):
\begin{align}\label{eq:mixed_bound}
  |\langle \logder_C\rangle|\leq 2\sqrt{\mathcal{I}_+\mathcal{I}_-}.
\end{align}
Unlike the other subordinate bounds, this speed limit is specific to the evolution of the system through state space.
We have thus found a family of uncertainty relations wherein the Fisher information (and its constituents) upper bound the timescales governing the evolution of tangent space observables.

Returning to our heuristic argument, we choose the instantaneous Lyapunov exponent, $r(t) := \langle \stability_+ \rangle$ as an observable to numerically confirm the validity of Eqs.~\ref{eq:tiur} and~\ref{eq:tiur_parts} for a sample chaotic orbit of the Lorenz model.
Figure~\ref{fig:speed_chaos}(a) shows the time evolution of the tangent space Fisher information (see Eq.~\ref{eq:def_FI}) setting the upper limit on the speed of $\langle \stability_+ \rangle$ (computed from Eq.~\ref{eq:def_tau}).
For $\stability_+$, $\speedtimep ^{-1}$ saturates the bound $\sqrt{\mathcal{I}_+}$, Fig.~\ref{fig:speed_chaos}(b).
The piece $\sqrt{\mathcal{I}_-}$ bounds the speed $\tau_{\stability_+,-}^{-1}$ in Fig.~\ref{fig:speed_chaos}(c) while the mixed term satisfies the inequality in Eq.~\ref{eq:mixed_bound} shown in Fig.~\ref{fig:speed_chaos}(d). The speed limits in Eqs.~\ref{eq:tiur} and \ref{eq:tiur_parts} saturate when the observable $\obs$ is $\stability^\top$ and $\stability_\pm$, respectively (SM VI).
To verify all of the the speed limits here, we examined them numerically for various observables and dynamics (SM): the Lorenz model in SM VIIB and the H\'enon-Heiles system in SM~VIIC.
We also analyzed the inverted harmonic potential to clarify the connection to Ruelle-Pollicott resonances~\cite{Gas2006}, SM VIIA.

\textit{Conclusions.---} Uncertainty relations are one of the most prominent features of quantum mechanics.
However, classical systems are also characterized by a type of uncertainty -- deterministic chaos -- in which the uncertainty in the initial conditions of dynamical systems devolves into chaotic behavior with characteristic timescales set by the Lyapunov exponents.
Here, we show that for a broad class of dynamical systems, this uncertainty and the sensitivity to initial conditions must obey (time-information) uncertainty relations.
These classical uncertainty relations are speed limits on the evolution of tangent space observables, including the instantaneous Lyapunov exponents whose time average are often studied phase space invariants and whose sum is the rate of phase space contraction that is connected to the entropy production in thermostatted dynamics~\cite{dorfmanIntroductionChaosNonequilibrium1999}.
These speed limits derive from a classical density matrix formulation of deterministic dynamical systems that has parallels with quantum mechanics, suggesting possibilities for further cross pollination of theories.
For example, by defining observables in the tangent space, we derived their equation of motion, which is similar to Ehrenfest's theorem.
Among these observables, it is the Fisher information for deterministic dynamical systems that appears in all of these time-information uncertainty relations, setting the classical speed limit, and partitioning into symmetric and anti-symmetric parts that set subordinate speed limits on classical observables.
All of these speed limits are model independent and transform the longstanding statistical nature of uncertainty relations into a mechanical picture, with potential for broad applications to engineered and living systems.\\

This material is based upon work supported by the National Science Foundation under Grant No. 2124510 and 1856250.
This publication was also made possible, in part, through the support of a grant from the John Templeton Foundation.

%

\clearpage

\setcounter{equation}{0}
\setcounter{figure}{0}
\renewcommand{\theequation}{SM\arabic{equation}}
\renewcommand{\thefigure}{SM\arabic{figure}}
\title{Supplemental Material: Speed limits on classical chaos}
\maketitle

\section{Equation of motion of the mean of an observable}

The mean of an arbitrary observable represented by a symmetric matrix $\obs$ for a general pure state $\brho$ is $\langle\obs\rangle = \Tr(\brho\,\obs)$.
Geometrically, the mean of an observable $\langle\obs\rangle = \Tr(\obs\brho)$ for a pure perturbation state is the scalar projection of $\ket{\delta\yu}$ on $\obs\ket{\delta\yu}$.
For instance, the instantaneous Lyapunov exponent is the projection of $\ket{\delta\yu}$ on $\stability_+\ket{\delta\yu}$.
The mean $\langle \stability_-\rangle=0$, so the anti-symmetric matrix $\stability_-$ continuously projects $\ket{\delta\yu}$ onto an orthogonal vector at each moment of time (i.e., $\stability_-$ rotates $\ket{\delta\yu}$ by $\pi/2$).
The time derivative of the mean is:
\begin{align}
d_t\langle\obs\rangle &= \Tr(\obs d_t\brho) + \Tr(\brho d_t\obs)\nonumber\\
&= \Tr(\obs\{\stability_+,\brho\}) +  \Tr(\obs[\stability_-,\brho]) -2\langle\stability_+\rangle\Tr(\brho\obs) \nonumber \\
& \quad + \Tr(\brho\,d_t\obs) \nonumber\\
& = 2(\langle \obs\,\stability_+\rangle -\langle\obs\rangle \langle\stability_+\rangle) + \langle [\obs,\stability_-]\rangle + \left\langle d_t\obs\right\rangle \nonumber\\
& = \text{cov}(\obs,\,2\stability_+) - \text{cov}(\obs,\,2\stability_-)+ \left\langle d_t\obs\right\rangle, \nonumber \nonumber\\
& = \text{cov}(\obs,2\stability^\top)+ \left\langle d_t\obs \right\rangle.\label{eq:EOM_mean_obsSM}
\end{align}
where $\text{cov}(\bm X,\bm Y) = \langle \bm X \bm Y^\top\rangle - \langle \bm X\rangle \langle \bm Y^\top\rangle$.  We have also used the following relations: $\langle\{\obs, \stability_-\} \rangle = 0$ and $\langle \stability_- \rangle = 0$.
For example, the time-derivative of the ILE is given by
\begin{align}
\dot{r} &= d_t\langle\stability_+\rangle =  2\Delta\stability_+^2 + 2\langle \stability_+\stability_- \rangle + \left\langle d_t\stability_+\right\rangle. \nonumber
\end{align}
For Hamiltonian dynamics, we can represent the relation using Poisson bracket $\{.\}_P$ as:
\begin{align}
\{r, H\}_P = 2\Delta\stability_+^2 + 2\langle \stability_+\stability_- \rangle+\left\langle d_t \stability_+\right\rangle. \nonumber
\end{align} 

\section{Equation of motion for observables in the density matrix basis}

Consider the time evolution equation of a mean observable $\langle\obs\rangle$  (Eq.~\ref{eq:EOM_mean_obsSM}). If the observable $\obs$ and $\brho$ commute, then they share the same set of eigenbasis. We then have
\begin{align}
\obs\brho = \brho\obs = \obs\ket{\delta \yu}\bra{\delta \yu} = \alpha\brho, \nonumber
\end{align}
where $\alpha$ is an eigenvalue of $\obs$.
The covariance term in Eq.~\ref{eq:EOM_mean_obsSM} becomes:
\begin{align}
\text{cov}(\obs,2\stability^\top) &= 2\langle \obs\stability \rangle -2\langle\obs\rangle \langle\stability\rangle \nonumber\\
& = 2\Tr(\obs\stability\brho) - 2\Tr(\obs\brho)\Tr(\stability\brho) \nonumber\\
& = 2\Tr(\brho\obs\stability) - 2\Tr(\obs\brho)\Tr(\stability\brho)  \nonumber \\
& = 2\alpha\Tr(\stability\brho) - 2\alpha\Tr(\stability\brho) = 0.\nonumber
\end{align}
This reduces Eq.~\ref{eq:EOM_mean_obsSM} to $d_t\langle\obs\rangle =  \left\langle d_t\obs\right\rangle$.

\section{Pure states}

\noindent Consider the equation of motion for $\brho$:
\begin{align}
d_t\brho = \bar{\stability}\brho + \brho\bar{\stability}^\top. \nonumber
\end{align}
We multiply $\brho$ form the left and right to write the following two equations:
\begin{align}
\brho\, (d_t \brho) &= \brho\bar{\stability}\brho + \brho^2\bar{\stability}^\top,\, \text{and} \quad (d_t\brho)\,\brho &= \bar{\stability}\brho^2 + \brho\bar{\stability}^\top\brho.\nonumber
\end{align}
Adding these equations,
\begin{align}
d_t(\brho^2) =  \bar{\stability}\brho^2 + \brho^2\bar{\stability}^\top\label{eq:der_rho},
\end{align}
where we have used $\brho\bar{\stability}\brho = \brho\bar{\stability}^\top\brho = 0.$
The right-hand side of Eq.~\ref{eq:der_rho} becomes $d_t \brho$ iff $\brho^2=\brho$. 

\section{Fisher information}

The symmetric and anti-symmetric parts of the stability matrix contributes to the Fisher information as follows (see the main text):
\begin{align}
\mathcal{I}_\pm = \langle\logder_\pm^2\rangle = 4\Delta\stability_\pm^2. \nonumber
\end{align}
For a pure perturbation state, these quantities are given by:
\begin{align}
\Delta\stability_+^2 & = \langle\stability_+^2\rangle - \langle\stability_+\rangle^2 \nonumber\\
&= \bra{\delta \yu}\stability_+\stability_+\ket{\delta \yu} - \bra{\delta \yu}\stability_+\ket{\delta \yu}^2 \nonumber \\
&= \|\stability_+\ket{\delta \yu}\|^2 - \cos^2\alpha\|\stability_+\ket{\delta \yu}\|^2 \nonumber\\
& = (1- \cos^2\alpha)\|\stability_+\ket{\delta \yu}\|^2 = \sin^2\alpha\|\stability_+\ket{\delta \yu}\|^2, \nonumber
\end{align}
where $\alpha$ is the angle between the vectors $\ket{\delta \yu}$ and $\logder_+\ket{\delta \yu}$ given by $\cos\alpha = \frac{\bra{\delta \yu}\logder_+\ket{\delta \yu}}{\|\logder_+\ket{\delta \yu}\|}$.
\noindent Next, we have
\begin{align}
\Delta \stability_-^2 & = \langle\stability_-^2\rangle = \bra{\delta \yu} \stability_-\stability_-^\top\ket{\delta \yu} = \|\stability_-\ket{\delta \yu}\|^2. \nonumber
\end{align}
The resulting vector $\stability_-\ket{\delta \yu}$ is perpendicular to $\ket{\delta \yu}$:
\begin{align}
\bra{\delta \yu}\stability_-\ket{\delta \yu} = \Tr(\stability_-\brho) = 0 \nonumber,
\end{align}
as the trace of the product of a symmetric and an anti-symmetric matrices vanishes.
Finally, these two Fisher information contributions are:
\begin{align}
\mathcal{I}_+ &= \|\mathcal{L}_+\ket{\delta \yu}\|^2= 4\sin^2\alpha\,\|\stability_+\ket{\delta \yu}\|^2, \nonumber\\
\mathcal{I}_- &= \|\mathcal{L}_-\ket{\delta \yu}\|^2 = 4\|\stability_-\ket{\delta \yu}\|^2. \nonumber
\end{align}
Fig.~\ref{fig:vectors_on_orbit} illustrates this geometric interpretation.

\begin{figure}[t]
	\vspace{2mm}
	\begin{tikzpicture}
	\draw[draw=gray,ultra thick]
	(-1,4.05) to [out=20,in=135](5,5);
	\draw [-latex, draw=gray,ultra thick](0.46, 4.69) -- (1.0,4.95);

	\draw[dashed,->,line width=0.5mm](2,5.4) -- (3.2,5.4+2.4);
	\node[align=left] at (2.12,6.95) {$\ket{\delta \yu}$};
	
	\draw[draw=red,->,line width=0.5mm](2.0,5.38) -- (4,3.5); %
	\node[align=left] at (4.5,4.0) {$\logder_-\ket{\delta \yu}$};
	\draw[draw=blue,->,line width=0.5mm](2.0,5.38) -- (6,7); %
	\node[align=left] at (4.8,6) {$\logder_+\ket{\delta \yu}$};
	
	\draw[draw=black,-,line width=0.2mm](2.08,5.53) -- (2.356, 5.3);  %
	\draw[draw=black,-,line width=0.2mm](2.35, 5.31) -- (2.26, 5.15); %
	
	\draw[draw=black,thin]
	(2.2,5.8) to [out=20,in=95](2.5,5.57); %
	\node[align=left] at (2.6,5.92) {$\alpha$};  %
	\draw[draw=black,thin]
	(2.7,5.66) to [out=-5,in=15](2.6,4.85); %
	\node[align=left] at (2.6,5.25) {$\gamma$};  %
	
	\node[align=left] at (4.6,7.25) {$\mathcal{I}_+ = \|\logder_+\ket{\delta \yu}\|^2$};
	\node[align=left] at (1.5,3.950) {$\mathcal{I}_- = \|\logder_-\ket{\delta \yu}\|^2$};
	\end{tikzpicture}
	
	\caption{A perturbation vector $\ket{\delta \yu}$ on a given trajectory in phase space. The vector $\stability_+\ket{\delta \yu}$ makes an angle $\alpha$ with $\ket{\delta \yu}$ and the vector $\logder_-\ket{\delta \yu}$ is orthogonal to $\ket{\delta \yu}$. The Fisher information contributors $\mathcal{I}_\pm$ are given by the magnitudes of these vectors as shown.\label{fig:vectors_on_orbit}}
\end{figure}

Next, consider the \emph{mixed} term $\logder_C = 4\,[\stability_+,\stability_-]$. Its expectation value can be expressed in several ways:
\begin{align}
\langle \logder_C\rangle &= \langle[\stability_+,\stability_-] \rangle = 8\,\text{cov}(\stability_+,\stability_-) \nonumber \\
&= 2(\Delta \stability^{\top2} - \Delta\stability^2) = 4(\Delta\stability_+^2+\Delta\stability_-^2-\Delta\stability^2). \nonumber
\end{align}
It then follows the cosine law for vectors $\logder^\top\ket{\delta \yu}$ and $\logder\pm\ket{\delta \yu}$ -- if $\gamma$ is the angle between the latter two vectors (Fig.~\ref{fig:vectors_on_orbit}), we get $\langle \logder_C\rangle = 8\Delta\stability_+\Delta\stability_-\cos\gamma$.

\section{Bound on \texorpdfstring{$\logder_C$}{TEXT}}

Consider the mean $\langle \logder_C \rangle$:
\begin{align}
\langle \logder_C \rangle &= 4\,\langle [\stability_+,\stability_-]\rangle  = 8\,\text{cov}(\stability_+,\stability_-). \nonumber
\end{align}
Using the Cauchy-Schwarz inequality, we get
\begin{align}
8|\text{cov}(\stability_+,\stability_-)| = |\langle \logder_C\rangle| \leq 8\,\Delta \stability_+\Delta\stability_-= 2\sqrt{\mathcal{I}_+}\sqrt{\mathcal{I}_-}. \nonumber
\end{align}
So we get the following series of inequality:
\begin{align}
|\langle \logder_C\rangle| \leq 2\sqrt{\mathcal{I}_+\mathcal{I}_-}. \nonumber 
\end{align}

\section{Saturation of bounds}

The speed limits in Eqs. 11 and 16\ in the main text saturate when the observable $\obs$ is $\stability^\top$ and $\stability_\pm$, respectively. In these cases, covariances simply reduce to variances (and half of the associated Fisher information).
\begin{enumerate}[label=(\alph*)]
	\item $\tau_{\mathcal{O}} \sqrt{\mathcal{I}_F} \geq 1$: Choosing the observable to be $\stability^\top$, we get:
	\begin{align}
	\tau_{\stability^\top}^{-1} &= \frac{|\text{cov}(\stability^\top,\logder)|}{\Delta \obs} = \frac{|\text{cov}(\stability^\top,2\stability^\top)|}{\Delta \stability^\top}\nonumber \\
	& = 2\Delta \stability^\top = \sqrt{\mathcal{I}_F}. \nonumber
	\end{align}
	\item $\tau_{\mathcal{O},\pm} \sqrt{\mathcal{I}_\pm} \geq 1$: Choosing the observable to be $\stability_\pm$, we get:
	\begin{align}
	\tau_{\stability_\pm,\pm}^{-1} &= \frac{|\text{cov}(\stability_\pm,\logder)|}{\Delta \obs} = \frac{|\text{cov}(\stability_\pm,2\stability^\top)|}{\Delta \stability^\top} \nonumber\\
	& = 2\Delta \stability_\pm = \sqrt{\mathcal{I}_\pm}. \nonumber
	\end{align}
\end{enumerate}
Also, notice that saturation of the bounds occurs when the observable $\obs$ ($\obs_{\pm}$) is linearly related to the logarithmic derivative $\logder$ ($\logder_\pm$).

\section{Speed limits on some observables}

\subsection{The inverted harmonic potential}

The inverted harmonic potential is a simple system to analyze the dynamical instability in transient and scattering processes. It is given by the Hamiltonian $H(x,\,y)= \lambda x y$, where $\lambda$ is a parameter that determines the Lyapunov exponent of the system. Its equations of motion are given by:
\begin{align}
\dot{x} = \lambda x, \quad \dot{y} = -\lambda y.
\end{align}
The inverse of $\lambda$ defines the exponential decay lifetime of an ensemble of trajectories: $\lambda = \tau_D^{-1}$. Its stability matrix,
\begin{align}
\stability_+ = \lambda\begin{pmatrix}
1 &	0\\ 
0 & -1
\end{pmatrix},
\end{align}
is symmetric.%

\begin{figure}[b!]
	\centering
	\includegraphics[width=0.75\columnwidth]{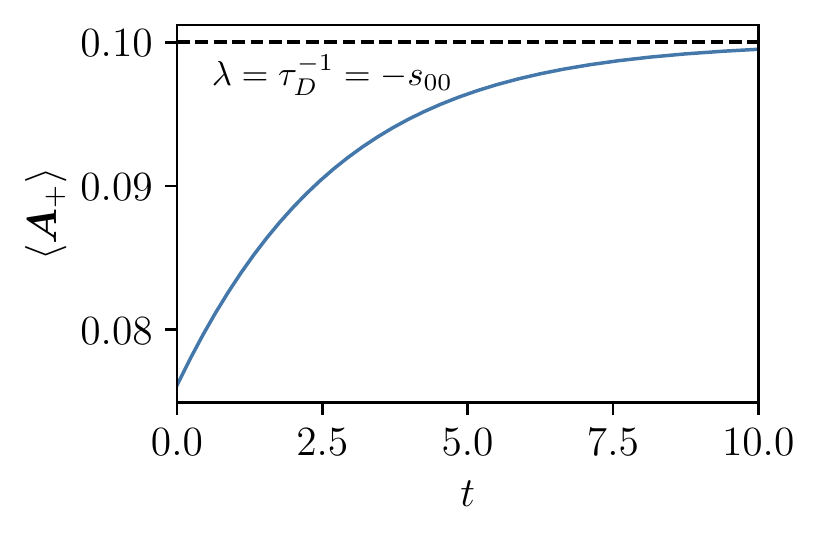}
	\vspace{-0.5cm}
	\caption{\label{fig:ILE_IHO}Instantaneous Lyapunov exponent for an arbitrary perturbation state in the inverted harmonic potential for $\lambda = 0.10$.}
\end{figure}
Gaspard~\cite{Gas2006a} showed that the decay rate $\tau_D$ is related to the generalized eigenvalues of the Liouvillian operator, otherwise known as the Ruelle-Pollicott (RP) resonances~\cite{Chaos_book_averaging}.
These resonances are given by integer multiples of the Lyapunov exponent $\lambda$, $s_{lm} = -\lambda(l + m +1) = -\tau_D^{-1}(l + m +1); \,l,m = 0,1,2,\cdots$.
The leading RP resonance (when $l=m=0$) sets the decay rate: $s_{00} = -\lambda = -\tau_D^{-1}$.
For a general perturbation state, 
\begin{align}
\brho = \frac{1}{\delta q^2 + \delta p^2}\begin{pmatrix}
\delta q^2 &	\delta q\delta p\\ 
\delta q \delta p & \delta p^2
\end{pmatrix}, 
\end{align}
the instantaneous Lyapunov exponent for the system is given by:
\begin{align}
r = \langle \stability_+\rangle = \lambda\left(\frac{\delta q^2 \,- \delta p^2}{\delta q^2\, +\delta p^2}\right) = \tau_D^{-1}\left(\frac{\delta q^2 \,- \delta p^2}{\delta q^2\, +\delta p^2}\right) .
\end{align}
Its upper and lower bounds are set by $\lambda$,
\begin{align}
-\lambda = s_{00} \leq \langle\stability_+\rangle \leq \lambda=\tau_D^{-1} = -s_{00}.
\end{align} 
We thus see that, for this system, the leading RP resonance which is an eigenvalue of $\stability_+$, provides the lower bound on instantaneous Lyapunov exponent.

The Fisher information of the system is also related to the leading RP resonance,
\begin{align}
\mathcal{I}_F = 4(\lambda^2 - r^2) = 4(s^2_{00} - r^2).
\end{align}
Since $\stability_+$ is the same as $\stability$, the intrinsic speed of the evolution of $\langle \stability_+\rangle$, $\tau_{\stability_+}^{-1}$ saturates $\mathcal{I}_F$.
\begin{figure}[h!]
	\centering
	\includegraphics[width=0.8\columnwidth]{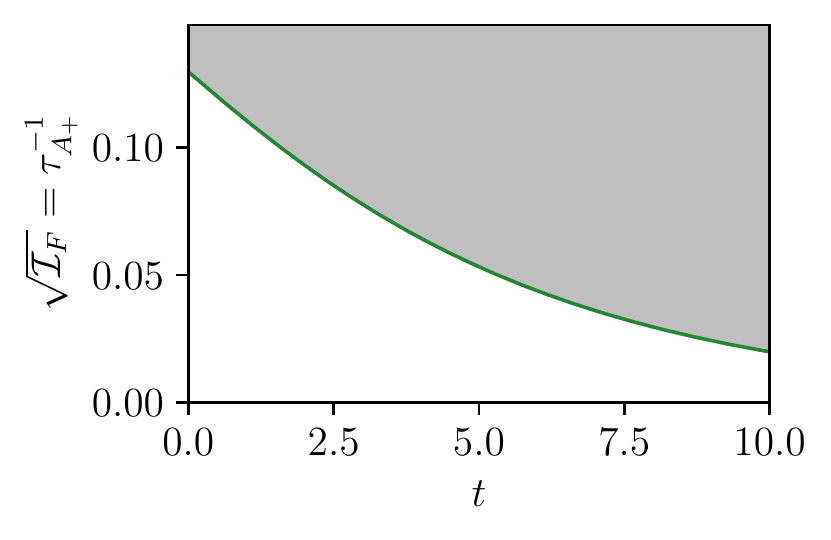}
	\caption{\label{fig:FI_IHO}Time evolution of $\tau_{\stability_+}^{-1}$ for the state used to compute instantaneous Lyapunov exponent in Fig.~\ref{fig:ILE_IHO}. The shaded region marks speeds not accessible to $\langle \stability_+\rangle$.}
\end{figure}

\subsection{The Lorenz model}

We consider the model of atmospheric convection due to Lorenz and Fetter~\cite{Lorenz63}. The model is defined by the following set of ordinary differential equations,
\begin{align}
\dot{x} &= \sigma(y-x), \quad \dot{y} &= x(\rho - z) -y,  \quad
\dot{z} &= xy - \beta z,\nonumber
\end{align}

\begin{figure}[t!]
	\vspace*{0.25cm}
	\centering
	The Lorenz model: Observable $\stability_-$\\
	\includegraphics[width=0.45\textwidth]{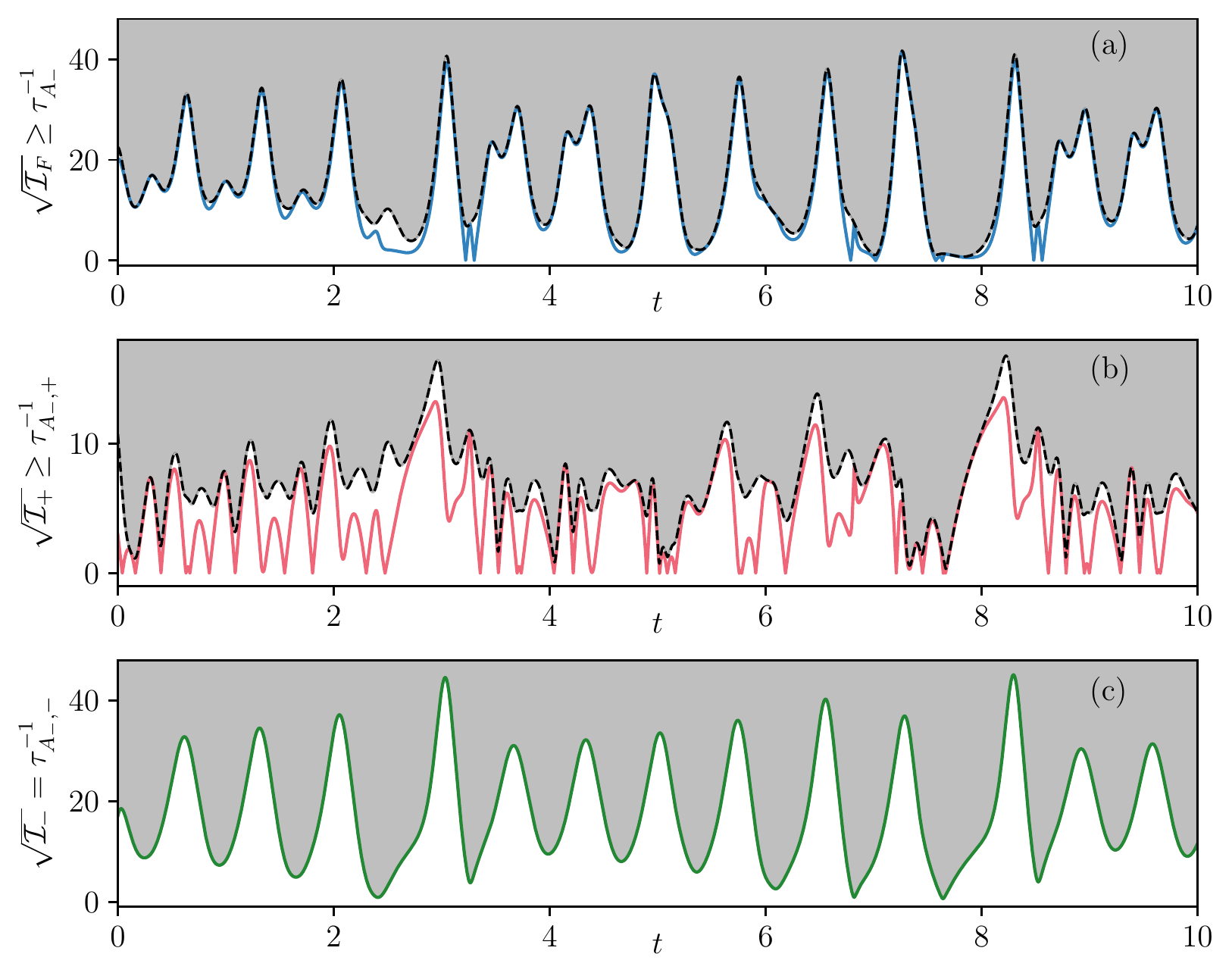}
	\caption{\label{fig:Lorenz2}The mean observable $\langle \stability_-\rangle$, for a chaotic orbit of the Lorenz model.--- (a) Square root of the tangent-space Fisher information (dashed black) $\sqrt{\mathcal{I}_F}$ upper bounds the speed $\tau_{\stability_+}^{-1}$ (solid blue). (b) The speed $\tau_{\stability_+,-}$ (solid red) is bounded by $\sqrt{\mathcal{I}_-}$ (dashed black). (c) For this observable, the speed $\tau_{\stability_+,+}^{-1}$ saturates the bound set by $\sqrt{\mathcal{I}_+}$ (solid green). Shaded regions mark speeds not accessible to the observable.}
\end{figure}

\begin{figure*}[b!]
	\centering
	The Lorenz model: Observable $\stability^\top$ \hspace{3cm} The Lorenz model: Observable $\stability$\\
	\includegraphics[width=0.45\textwidth]{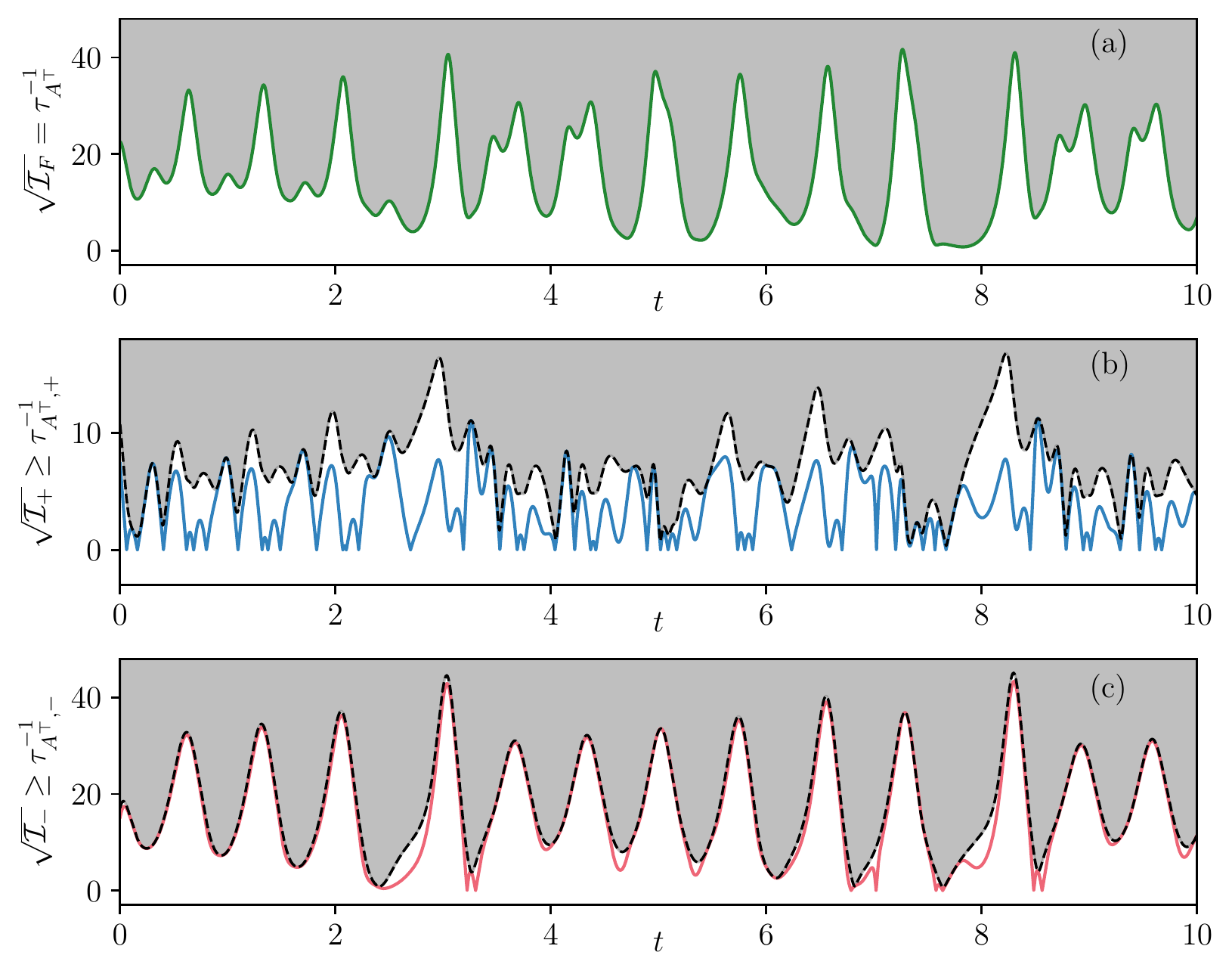}
	\includegraphics[width=0.45\textwidth]{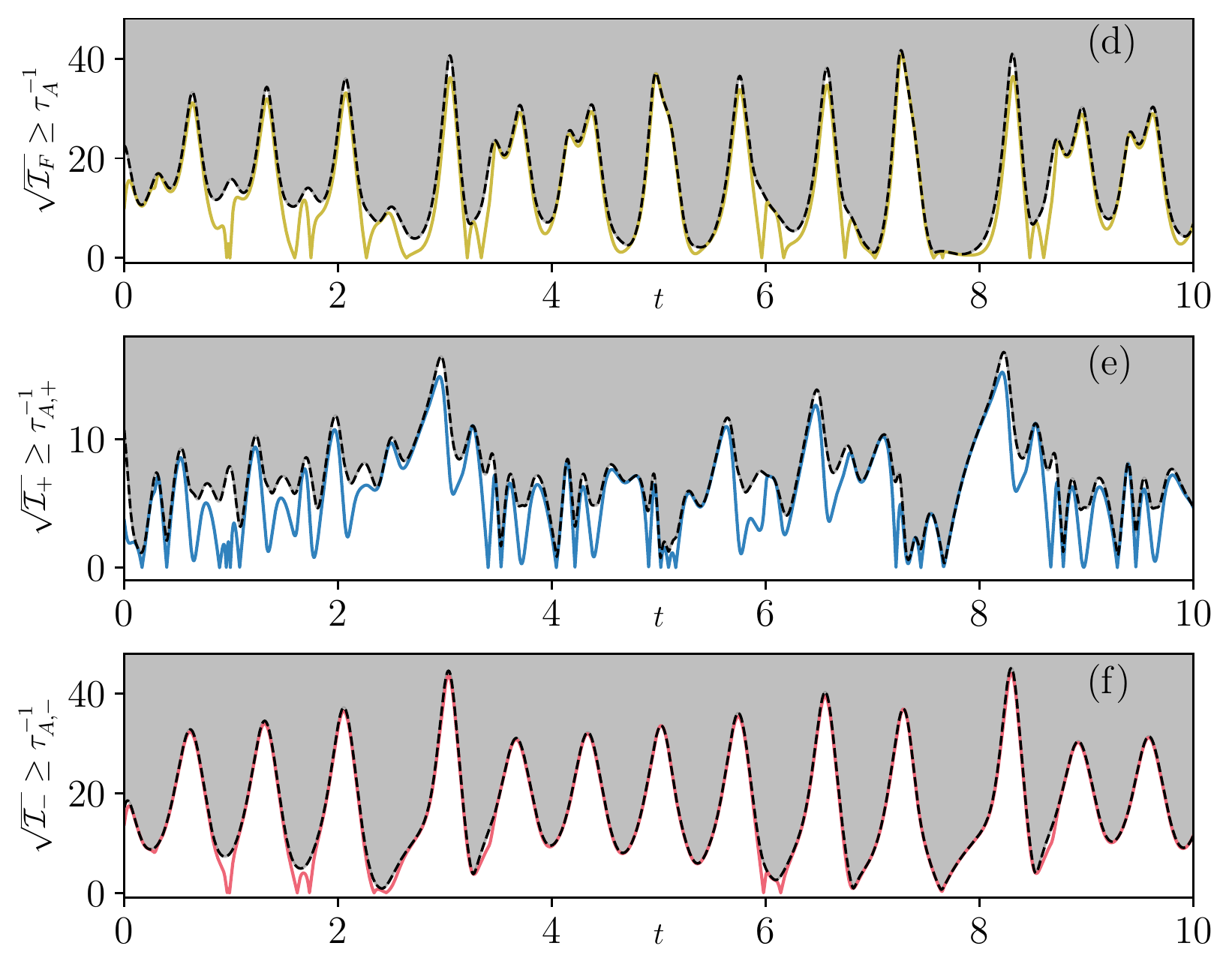}
	\caption{\label{fig:Lorenz3}Mean observables $\langle \stability^\top\rangle$ and $\langle \stability\rangle$, for a chaotic orbit of the Lorenz model.--- (a) the speed $\tau_{\stability^\top}^{-1}$ (solid green) saturates the square root of the tangent-space Fisher information (dashed black) $\sqrt{\mathcal{I}_F}$; $\sqrt{\mathcal{I}_+}$ (dashed black) upper bounds the speeds (solid blue) $\tau_{\stability^\top,+}^{-1}$  in (b) and $\tau_{\stability,+}^{-1}$ in (e); $\sqrt{\mathcal{I}_-}$ (dashed black) upper bounds the speeds (solid red) $\tau_{\stability^\top,-}^{-1}$ in (c) and $\tau_{\stability,-}^{-1}$ in (f); $\sqrt{\mathcal{I}_F}$ (dashed black) upper bounds the speed $\tau_{\stability}^{-1}$ (solid yellow) in (d). Shaded regions mark speeds not accessible to the observable.}
\end{figure*}

with the stability matrix:
\begin{align}
\stability =\begin{pmatrix}
-\sigma & \sigma & 0\\
\rho-z & -1 & -x\\
y & x & -\beta \nonumber
\end{pmatrix}.
\end{align}
Speed limits on mean observables $\stability_-,\stability^\top$ and $\stability$ for a chaotic orbit of the Lorenz model (parameters: $\sigma = 10$, $\beta = 8/3$, $\rho = 28$) are plotted in Figs.~\ref{fig:Lorenz2} and \ref{fig:Lorenz3}.

\subsection{The H\'enon-Heiles system}

\begin{figure*}
	\centering
	\begin{minipage}[t]{0.45\textwidth}
		The H\'enon-Heiles system: Observable $\stability_+$ \\
		\includegraphics[width=0.9\textwidth]{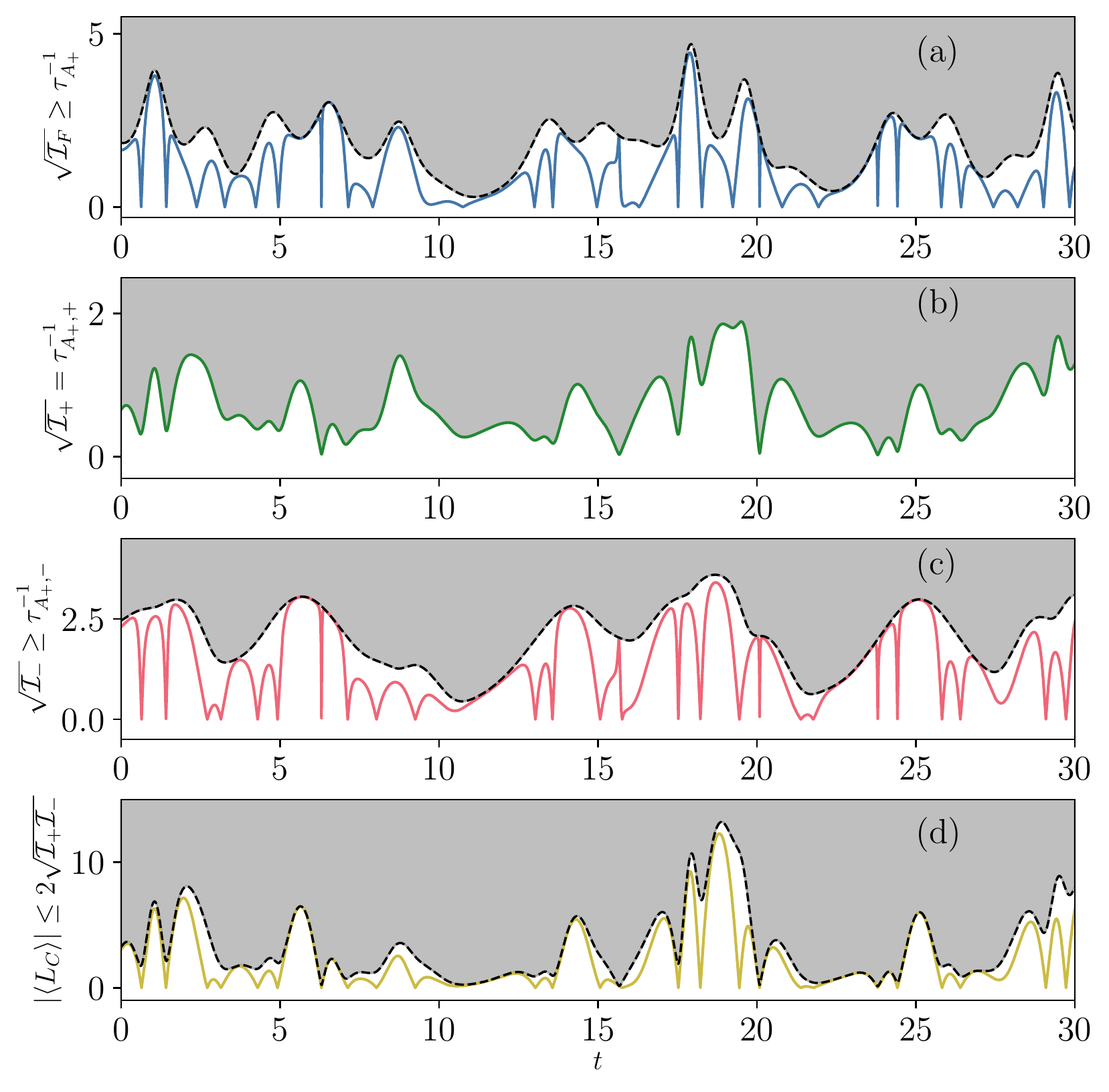}
		\caption{\label{fig:Henon1}\textit{Speed limit on chaos: the instantaneous Lyapunov exponent, $\langle \stability_+\rangle$, for a chaotic orbit of the H\'enon-Heiles system.---} (a) Square root of the tangent-space Fisher information (dashed black) $\sqrt{\mathcal{I}_F}$ upper bounds the speed $\tau_{\stability_+}^{-1}$ (solid blue). (b) For this observable, the speed $\tau_{\stability_+,+}^{-1}$ saturates the bound set by $\sqrt{\mathcal{I}_+}$ (solid green). (c) The speed $\tau^{-1}_{\stability_+,-}$ (solid red) is bounded by $\sqrt{\mathcal{I}_-}$ (dashed black). (d) The mixed term $|\langle\logder_C\rangle|$ (solid yellow) is bounded by 2$\sqrt{\mathcal{I}_+\mathcal{I}_+}$ (dashed black) from above. Shaded regions mark speeds not accessible to the observable.}
	\end{minipage}\qquad
	\begin{minipage}[t]{.45\textwidth}
		
		The H\'enon-Heiles system: Observable $\stability_-$
		\includegraphics[width=0.95\textwidth]{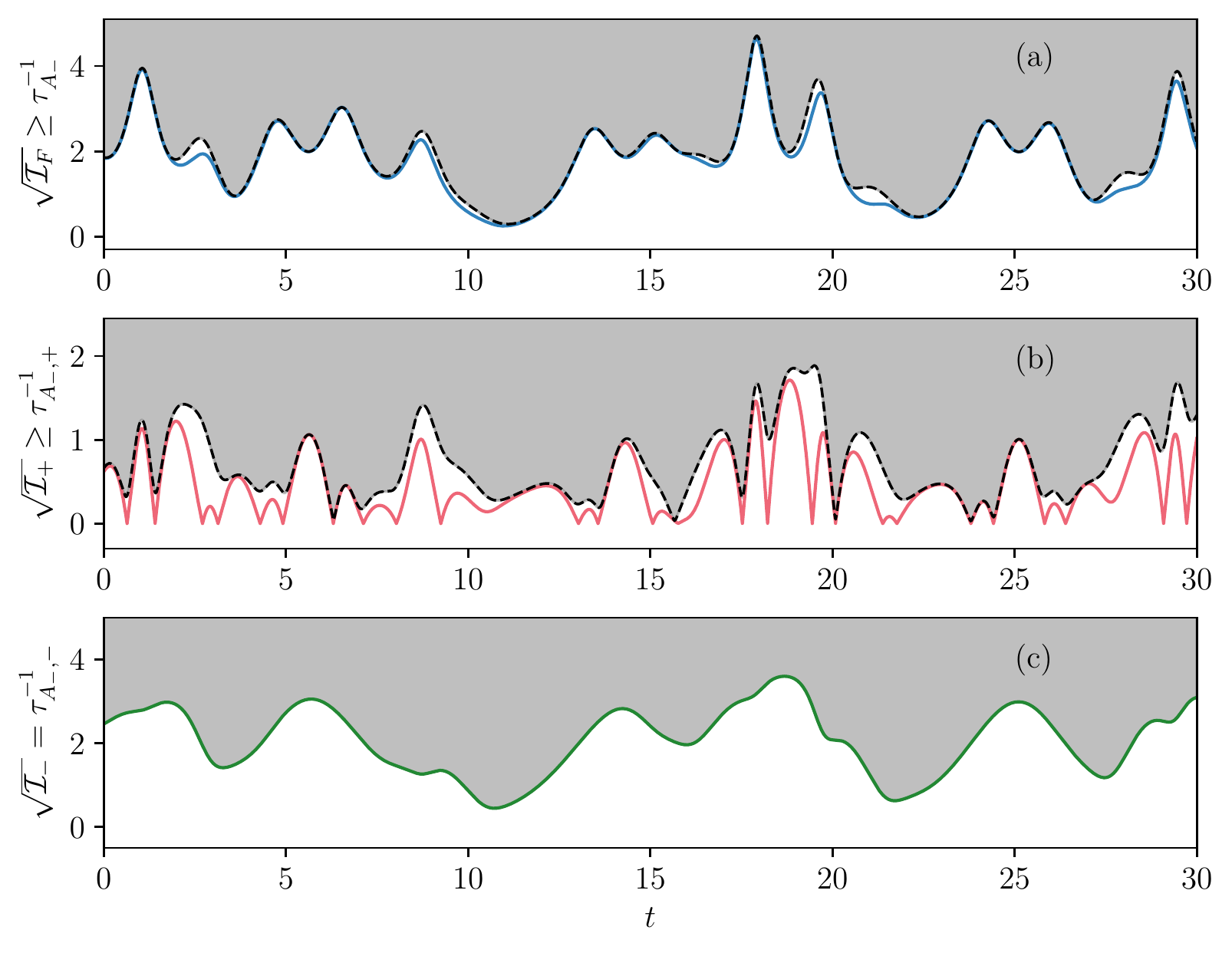}
		\vspace*{0.5cm}
		\caption{\label{fig:Henon2} The mean observables $\langle \stability_-\rangle$, for a chaotic orbit of the H\'enon-Heiles system.--- Bounding curves indicating Fisher information and its pieces are in dashed black. (a) Observable $\stability_-$ speed $\tau^{-1}_{\stability_-}$ (solid blue) is bounded by $\sqrt{\mathcal{I}_F}$. (b) The speed $\tau^{-1}_{\stability_-,+}$ (solid red) is bounded by $\sqrt{\mathcal{I_+}}$. (c)  The speed $\tau^{-1}_{\stability_-,-}$ (solid green) saturates the bound $\sqrt{\mathcal{I_-}}$. Shaded regions mark speeds not accessible to the observable.}
	\end{minipage}
\end{figure*}

\begin{figure*}[h!]
	\centering
	\vspace{1cm}
	The H\'enon-Heiles system: Observable $\stability^\top$ \hspace{2cm} The H\'enon-Heiles system: Observable $\stability$\\
	\includegraphics[width=0.48\textwidth]{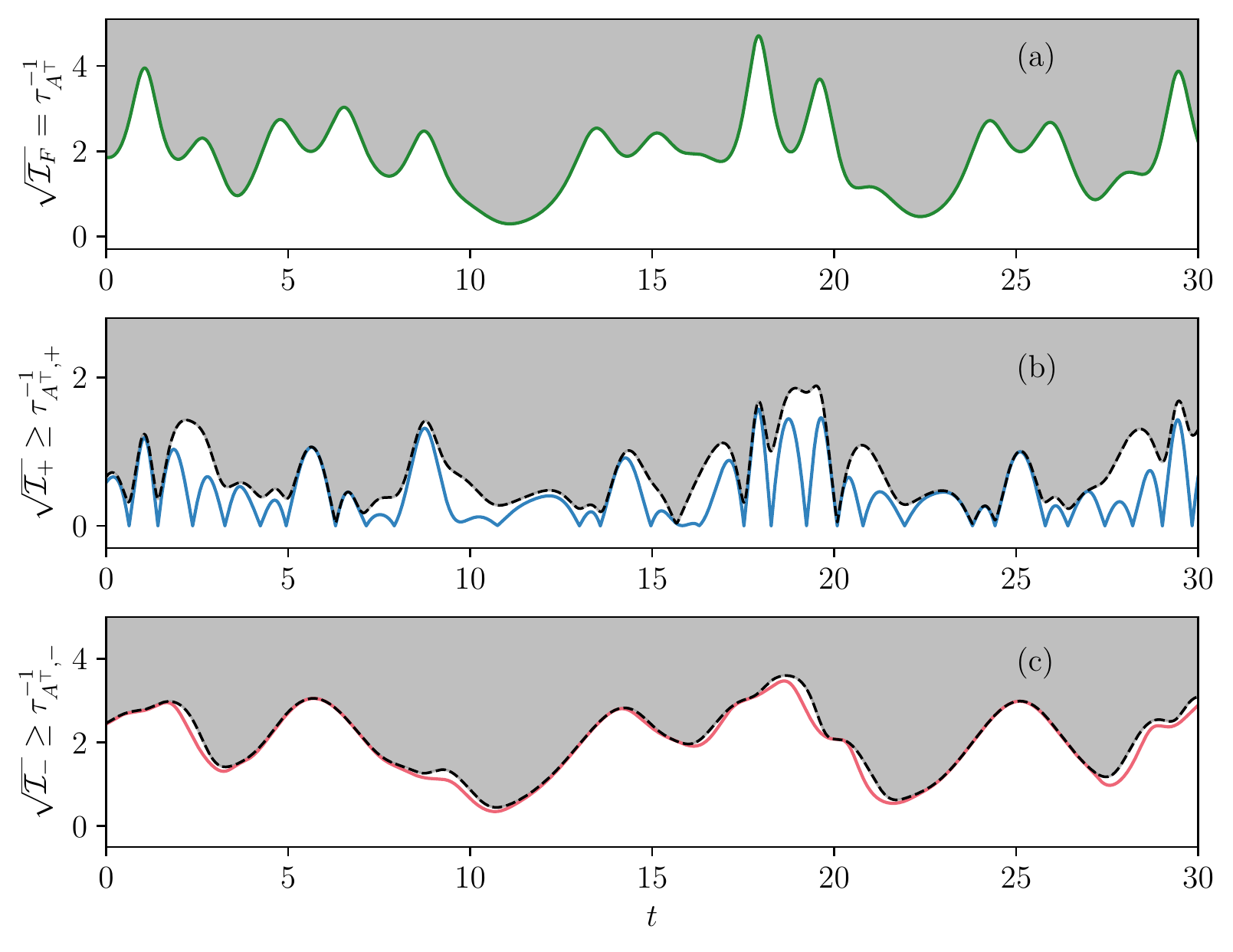}
	\includegraphics[width=0.48\textwidth]{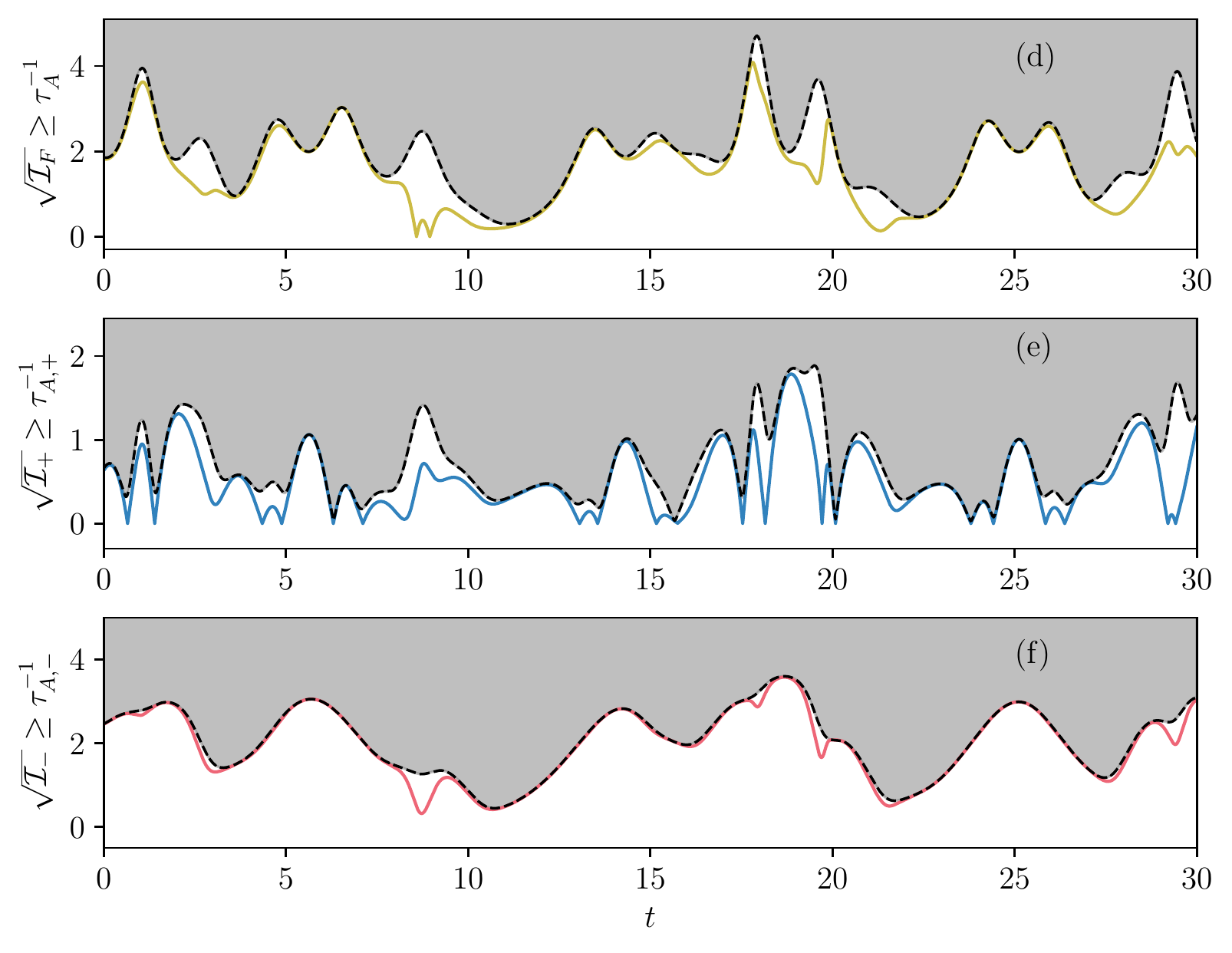}
	\caption{\label{fig:Henon3}Mean observables $\langle \stability^\top\rangle$, and $\langle \stability\rangle$, for a chaotic orbit of the H\'enon-Heiles system.--- Bounding curves indicating Fisher information and its pieces are in dashed black. (a) Observable speed $\tau^{-1}_{\stability^\top}$ (solid green) saturates the bound $\sqrt{\mathcal{I}_F}$. (b) The speed $\tau^{-1}_{\stability^\top,+}$ (solid blue) is bounded by $\sqrt{\mathcal{I}_+}$. (c) The speed $\tau^{-1}_{\stability^\top_-}$ (solid red) is bounded by $\sqrt{\mathcal{I}_-}$. (d) Observable speed $\tau^{-1}_{\stability}$ (solid yellow) is bounded by $\sqrt{\mathcal{I}_F}$. (e) The speed $\tau^{-1}_{\stability,+}$ (solid blue) is bounded by $\sqrt{\mathcal{I}_+}$. (f) The speed $\tau^{-1}_{\stability,-}$ (solid red) is bounded by $\sqrt{\mathcal{I}_-}$. Shaded regions mark speeds not accessible to the observable.}
\end{figure*}

We consider the H\'enon-Heiles model~\cite{Henon1964} to numerically verify the speed limits obtained in the main text. 
It system is given by the following system of equations,
\begin{align}
\dot{x} &= p_x, \quad \dot{y} = p_y,\nonumber\\
\dot{p}_x &= -x - 2xy, \quad \dot{p}_y  = -y - x^2 + y^2,\nonumber
\end{align}
which lead to the stability matrix:	
\begin{align}
\stability =
\begin{pmatrix}
0 & 0 & 1 & 0\\
0 & 0 & 0 & 1\\
-1 - 2y & -2x & 0 & 0\\
-2x & -1+2y & 0 & 0\nonumber
\end{pmatrix}.
\end{align}
For a chaotic orbit of the system corresponding to energy $E = 0.1666$, we obtain  plots in Figs. \ref{fig:Henon1}, \ref{fig:Henon2} and \ref{fig:Henon3} for bounds on speed of mean observables $\stability_\pm,\stability^\top$ and $\stability$.
\end{document}